\documentclass[aps,prl,twocolumn,superscriptaddress,nofootinbib]{revtex4}
\usepackage{graphicx}
\usepackage{bm}
\usepackage{tabularx}
\usepackage{epsfig}
\usepackage{amsfonts}
\usepackage{amsmath}

\usepackage{ upgreek }
\usepackage{color}

\usepackage{color}

\newcommand{\la}{\label}

\newcommand{\bbm}{\begin{multline}}
\newcommand{\eem}{\end{multline}}
\newcommand{\be}{\begin{equation}}
\newcommand{\ee}{\end{equation}}
\newcommand{\bea}{\begin{eqnarray}}
\newcommand{\eea}{\end{eqnarray}}
\newcommand{\p}{\partial}

\newcommand{\comment}[1]{}
\newcommand{\bsym}{\boldsymbol}

\begin{document}


\title{\large{Magnetotransport in Dirac metals: chiral magnetic effect and quantum oscillations}}

\author{Gustavo M.~Monteiro}
\affiliation{Department of Physics and Astronomy, Stony Brook University, Stony Brook, NY 11794, USA}

\author{Alexander G.~Abanov}
\affiliation{Department of Physics and Astronomy, Stony Brook University,  Stony Brook, NY 11794, USA}
\affiliation{Simons Center for Geometry and Physics,
Stony Brook University,  Stony Brook, NY 11794, USA}

\author{Dmitri E.~Kharzeev}
\affiliation{Department of Physics and Astronomy, Stony Brook University, Stony Brook, NY 11794, USA}
\affiliation{Department of Physics, Brookhaven National Laboratory, Upton, New York 11973-5000, USA}
\affiliation{RIKEN-BNL Research Center, Upton, New York 11973-5000, USA}

\begin{abstract}
Dirac metals are characterized by the linear dispersion of fermionic quasi-particles, with the Dirac point hidden inside a Fermi surface. We study the magnetotransport in these materials using chiral kinetic theory to describe within the same framework both the negative magnetoresistance caused by  chiral magnetic effect and quantum oscillations in the magnetoresistance due to the existence of the Fermi surface. We discuss the relevance of obtained results to recent measurements on ${\rm Cd_3As_2}$.
\end{abstract}

\maketitle

The discovery of Dirac semimetals \cite{Wang2012,Wang2013,Borisenko2014,Neupane2014,Liu2014} has enabled the experimental studies of 3-dimensional materials with chiral quasiparticles. In comparison to 2-dimensional graphene, the access to three spatial dimensions allows one to study phenomena such as chiral anomaly \cite{Adler1969,Bell1969, nielsen1983adler} and chiral magnetic effect (CME) \cite{CME2008}. In particular, the latter refers to the generation of electric current induced by chirality imbalance in the presence of the magnetic field; see \cite{kharzeev2014chiral,Burkov:2015hba} for reviews. 

Dirac semimetals together with Weyl semimetals are representatives of three-dimensional chiral materials. They are both characterized by the existence of band-touching points. In Weyl semimetals, the time reversal symmetry is spontaneously broken and each Dirac node splits into two disjoint Weyl points. In a Dirac semimetal, each Dirac node has zero Chern number, reflecting the coexistence of two Weyl points of opposite chiralities at the same point in the Brillouin zone. This corresponds to an emergent $\mathbb Z_2$-symmetry relating the two states with different chiralities. In parallel electric and magnetic fields, the degeneracy between the states with opposite chiralities gets broken due to the chiral anomaly and the difference between the Fermi energies of left- and right-handed fermions can be described by the chiral chemical potential $\mu_5$. This difference generates a nonvanishing chiral magnetic current of the form $\bsym j_{CME}=e^2\mu_5\bsym B/(2\pi^2)$.

Because of the chiral anomaly, $\mu_5 \sim \bsym{E}\cdot\bsym{B}$. Consequently, the CME conductivity acquires a positive term proportional to $B^2$ and the magnetoresistance (MR) becomes negative \cite{SonSpivak2013}. Such a behavior signaling the presence of CME has been observed recently in Dirac semimetals ZrTe$_5$ \cite{Kharzeev-ZrTe_5} and  Na$_3$Bi \cite{Na_3Bi-Ong-2015}. Negative MR has also been observed \cite{TaAs-Weyl-2015, TaAs-CME-2015} in TaAs, a candidate for a Weyl semimetal. On the other hand, the previous studies of magnetotransport in another candidate for a Dirac semimetal Cd$_3$As$_2$ \cite{Ong2014, Analytis-2015} revealed a more complicated pattern, with strong oscillations of MR.

Quantum oscillations in MR signal the presence of large Fermi surface. The material represents a Dirac metal rather than a semimetal with the Dirac point hidden inside a Fermi surface.  
In this Letter we develop the theory of magnetotransport in Dirac metals that describes an interplay between the CME and quantum Shubnikov-de Haas (SdH) oscillations. Our analysis can be trivially extended to Weyl metals if one assumes only one chirality per Dirac node.

Although Dirac metals are characterized by linear dispersion of quasiparticles $\varepsilon(\bsym k)=\hbar v_F |\bsym k|$\footnote{We assumed the Dirac point to be at $\bsym k=\bsym0$.}, the assumption of linear spectrum is absent in this Letter. However, we assume that the quasiparticles can be described by the Fermi liquid theory. Also, we restrict ourselves to isotropic system. The chemical potential or Fermi energy define the size of Fermi surface $\varepsilon_F=\varepsilon (k_F)$, where the Fermi momentum is related to the density of conduction electrons (per Dirac point and chirality) by standard formula $k_F=(3\pi^2 n_e)^{1/3}$.

\begin{figure}[h]
    \centering
    \includegraphics[width=0.47\textwidth]{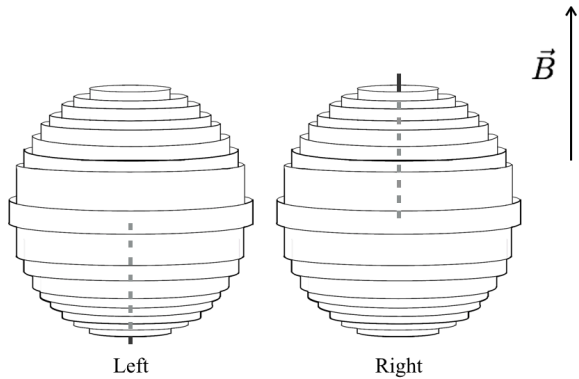}
    \caption{Semiclassical picture of Fermi surface for right and left chiral modes is shown in the presence of the magnetic field. For the linear dispersion, the support of Hamiltonian eigenstates is a collection of cylinders corresponding to eigenvalues $\varepsilon_{n}(k_z)=\hbar v_F\sqrt{k_z^2+k_\perp^2}$, where $k_\perp^2=2eB n$ with $n=0,1,2,\ldots$. The state with $n=0$ is chiral and exists only for $p_z>0$ (parallel to $B$) for the right chirality and for $p_z<0$ for the left one.}
    \label{fig:FermiSurface}
\end{figure}

Our treatment will be based on the semiclassical approximation, valid when magnetic field is weak enough so that a large number of Landau levels is filled. Introducing the magnetic length $\ell_B\equiv\sqrt{\frac{\hbar}{eB}}$ this condition amounts to $\frac{1}{2}k_F^2\ell_B^2\gg 1$. In this limit, we can associate a trajectory to each electron quasiparticle in the vicinity of the Fermi surface.\footnote{In this limit one can still think about Fermi sphere albeit stratified into Landau level ``cylinders'', see Figure~\ref{fig:FermiSurface}.} We expect to see pronounced SdH oscillations when the temperature is much smaller than the energy gap between Landau levels or, equivalently, when the thermal de Broglie wavelength is much bigger than the Larmor radius $\frac{\hbar v_F}{T}\gg k_F\ell_B^2$.\footnote{We measure temperature in energy units.}

At low temperatures, the impurity scattering is the leading contribution to conductivity tensor. This introduces another scale into the problem -- the scattering rate. The system is called clean when the quasiparticle performs many cyclotron orbits before colliding or, equivalently, when the mean-free-path is much larger than the cyclotron radius, $v_F\tau\gg k_F\ell_B^2$. We restrict ourselves to single-impurity scattering approximation and neglect interference and localization. This approximation is valid when the density of impurities is low.

Thus the regime of interest in this work is defined by
\bea
	1\ll (k_F \ell_B)^2 \ll k_F v_F \tau\,,
 \la{scales}
\eea
where for the linear spectrum the last term correspond to $\varepsilon_F\tau/\hbar$.

Dirac points in 3 dimensions correspond locally to monopole solutions of the Berry curvature in momentum space \cite{Volovik:2003fe}, which give rise to chiral anomaly effects in kinetic theory \cite{stephanov2012chiral, son2012berry}; see \cite{Basar:2013iaa} for application to Weyl semimetals. The Berry curvature is obtained from the Bloch functions for the valence band electron quasiparticles $|u_{\boldsymbol k}\rangle$ and  for an isotropic system is given by \cite{Bernevig-TI-TS}:
\be
	\bsym \Omega(\bsym k)
	= i\boldsymbol\nabla_{\boldsymbol k}\times\langle u_{\boldsymbol k}|\boldsymbol
	\nabla_{\boldsymbol k}u_{\boldsymbol k}\rangle
	=\chi\frac{\boldsymbol{\hat k}}{2k^2}\,.
 \la{BCurv}
\ee

The Berry curvature (\ref{BCurv}) has an opposite sign for chiralities $\chi=\pm 1$ and $\boldsymbol{\hat k}$ denotes a unit vector in the direction of $\boldsymbol{k}$.
We refer to the modifications of kinetic theory by Berry curvature as to the chiral kinetic theory \cite{stephanov2012chiral, son2012berry}. In particular, the Berry curvature modifies the expression for the current density. Given a dispersion relation $\varepsilon(\bsym k)$, we introduce the group velocity vector $\bsym v_{\bsym k}=\frac{1}{\hbar}\bsym\nabla_{\bsym k} \varepsilon(\bsym k)$, so that the current density can be expressed as \cite{ChangNiu-1995, ChangNiu-1996, Wave-packet-Niu1999, XiaoChangNiu-Review}:
\begin{equation}
	\bsym j=-2e\int\limits_{\mathbb{BZ}}f
	\left[\bsym v_{\bsym k}
	+\frac{e}{\hbar}\left(\bsym v_{\bsym k}\cdot\bsym\Omega\right)\bsym B
	+\frac{e}{\hbar}\bsym E\times\bsym\Omega\right]\frac{\text d^3k}{(2\pi)^3}\,,
 \la{current}
\end{equation}
where $f(\bsym x,\bsym k, t)$ is the distribution function and the integral is performed over the first Brillouin zone ($\mathbb{BZ}$). The overall factor of 2 accounts for spin projections. 

The distribution function is obtained by solving the Boltzmann equation. Since we are interested in linear response, we expand $f(\bsym x,\bsym k, t)$ around the equilibrium (Fermi-Dirac) distribution function $f_0(\varepsilon)$:
\begin{equation}
	f(\bsym x,\bsym k, t)=f_0(\varepsilon)
	+e\frac{\p f_0}{\p\varepsilon}\boldsymbol E\cdot\boldsymbol g 
	+\mathcal O(\boldsymbol E^2)\,. 
 \la{near-eq}
\end{equation}

Having $\boldsymbol g(\bsym x,\bsym k,t)$ from the linearized Boltzmann equation (see below) and substituting the ansatz (\ref{near-eq}) into (\ref{current}), the conductivity tensor reads:
\begin{align}
	\upsigma_{ab}=\;&-2e^2\int
	\frac{\p f_0}{\p\varepsilon}g_b\left(\bsym v_{\bsym k}
	+\frac{e}{\hbar}\left(\bsym v_{\bsym k}\cdot\bsym\Omega\right) 
	\bsym B\right)_{a}\frac{\text d^3k}{(2\pi)^3}\;+
 \nonumber \\
	&+\frac{2e^2}{\hbar}\,\varepsilon_{abc}\int
	\Omega_c(\boldsymbol k)f_0(\varepsilon)\,
	\frac{\text d^3k}{(2\pi)^3}\,,
 \la{conduc}
 \\
 =\;&-2 e^2\sum_{\chi=\pm} \int
	\frac{\p f_0}{\p\varepsilon} g_b\,v_k\left(\hat{\bsym k}
	+\chi\zeta_k\hat{\bsym z}\right)_{a}\frac{\text d^3k}{(2\pi)^3} \,.
 \la{conducsimpl}
 	\end{align}

Here and in the following all the expressions refer to a single Dirac node. The assumption that the Dirac points can be treated independently is valid when they are far apart in the Brillouin zone\footnote{In comparison to the Fermi momentum of each disjoint piece of Fermi surface.}, so that the quasiparticle scattering from one Dirac cone to the other requires a large momentum transfer. The last term in Eq. (\ref{conduc}) vanishes for isotropic dispersion relations. The last equality is obtained with the use of (\ref{BCurv}) assuming that the system is isotropic and that the integral is dominated by a vicinity to the Fermi surface due to the factor $\p f_0/\p \varepsilon$. We also introduced the small parameter (cf. (\ref{scales})) $\zeta_k = 1/(2k^2\ell_B^2)$ and considered that the magnetic field is along the $z$-direction. 

For Dirac metals, the $\mathbb Z_2$-symmetry holds at low energies and interaction terms that break this symmetry are sub-leading in comparison to the chirality-preserving ones. In this limit, the Boltzmann equations for different chiralities decouple and the collision integral accounts only for intra-chirality scattering. Given the transition rate $w_{\boldsymbol k'\rightarrow \boldsymbol k}$  from an initial state $\boldsymbol k'$ to a final state $\boldsymbol k$, the collision integral can be written as:
\begin{align}
	\mathcal I[f]=\int_{\mathbb{BZ}} 
	\left[f(\boldsymbol k')-f(\boldsymbol k)\right]
	w_{\boldsymbol k'\rightarrow \boldsymbol k}\; \Upsilon'	
	\frac{\text d^3k'}{(2\pi)^3} \,. 
 \la{col-int}
\end{align}

Here we assumed the elastic scattering probability to be invariant under time reversal, i.e. $w_{\boldsymbol k'\rightarrow \boldsymbol k}=w_{\boldsymbol k\rightarrow \boldsymbol k'}$. In addition to that, we have used $\left[f'\left(1-f\right)-f\left(1-f'\right)\right] = f(\boldsymbol k')-f(\boldsymbol k)$ and denoted the modification of the phase-space volume element due to a non-vanishing Berry curvature as
\be
	\Upsilon =1+\frac{e}{\hbar}\boldsymbol B\cdot\boldsymbol\Omega(\boldsymbol k)
	=1+\chi\zeta_k\cos\theta \,,
 \la{upsilon}
\ee
%
where $\theta$ is the angle between $\boldsymbol k$ and magnetic field.

Using the equations of motions from \cite{ChangNiu-1995, ChangNiu-1996, Wave-packet-Niu1999, XiaoChangNiu-Review}, the Boltzmann equation for $\bsym g(t,\bsym k)$ in the linearized regime becomes:
\begin{align}
	&\left[\Upsilon(\p_t+i\omega)
	-\frac{e}{\hbar}(\bsym v_{\bsym k}\times\bsym B)
	\cdot\boldsymbol\nabla_{\boldsymbol k}\right]\boldsymbol g
	=
 \la{Boltz-g} \\
	&=\bsym v_{\bsym k}
	+\frac{e}{\hbar}(\bsym v_{\bsym k}
	\cdot\bsym\Omega)\boldsymbol B +\int\limits_{\mathbb{BZ}}\frac{\text d^3k'}{(2\pi)^3}\;
	(\Upsilon' w_{\boldsymbol k'\rightarrow \boldsymbol k}\Upsilon)\,
	[\boldsymbol g'-\boldsymbol g]\,. 
 \nonumber
\end{align}

In Eq. (\ref{Boltz-g}), we have assumed that the system is uniform and the electric field oscillates with the frequency $\omega$, i.e., $\boldsymbol E=\boldsymbol E_0\, e^{i\omega t}$. It is straightforward to observe that this equation does not admit a stationary solution when $\omega=0$\footnote{This can be seen by integrating (\ref{Boltz-g}) over the solid angle.}. This is the manifestation of the chiral anomaly in kinetic theory -- the constant parallel electric and magnetic field continue to pump chirality into the system. However, a stationary solution does exist in the presence of a chirality relaxation mechanism.
 
To determine $w_{\boldsymbol k'\rightarrow \boldsymbol k}$, we assume that the elastic scattering occurs on weak, dilute, and point-like impurities.  We thus model the single-impurity scattering by
\begin{equation}
	w_{\boldsymbol k'\rightarrow \boldsymbol k}
	=\frac{3}{2\nu(\varepsilon)\tau(\varepsilon)}
	(1+\boldsymbol{\hat k'}\cdot \boldsymbol{\hat k})\,\delta(\varepsilon-\varepsilon')\,, 
 \la{scat-rate}
\end{equation}
where $\nu(\varepsilon)$ is the density of states -- in the absence of magnetic field -- at the energy $\varepsilon$. We assumed that the scattering is elastic and averaged over impurity positions. All microscopic details are absorbed into the transport scattering time $\tau$. We remark here that although we focused on the small wave vector limit, the scattering rate from Eq. (\ref{scat-rate}) is not isotropic. This is because the Weyl-particle spins are always polarized along their momenta, producing a universal factor $(1+\boldsymbol{\hat k'}\cdot \boldsymbol{\hat k})$, which suppresses the backscattering of particles by impurities. For example, for massless Dirac quasiparticles one can find at leading order in the partial-wave expansion of scattering amplitude\footnote{Although the magnetic field breaks the 3D rotation invariance, the assumption of adiabatic evolution allows us to write the eigenbasis in terms of Bloch functions or plane waves. The effect of magnetic field is absorbed into the trajectory in k-space and in the measure. A solution of the Dirac scattering problem can be found, e.g., in \cite{LandauLifshitz-4} and gives for scattering amplitude $
	A(\boldsymbol{\hat k'}\cdot \boldsymbol{\hat k})=\frac{\hbar v_F}{2i\varepsilon}
	\sum_{l=1}^{\infty}l\left(e^{2i\delta_l}-1\right)\left[P_l(\boldsymbol{\hat k'}\cdot 
	\boldsymbol{\hat k})+P_{l-1}(\boldsymbol{\hat k'}\cdot \boldsymbol{\hat k})\right]\,.
$.}:
$$
	\frac{1}{\tau}= n_{imp} \frac{2v_F}{3\pi^2k^2}\sin^2\delta_1 \,.
$$

The scattering phase $\delta_1$ in the general case should also depend on the magnitude of magnetic field since the screening of the impurity potential might be modified by $B$. Since $\bsym B=B\bsym{\hat z}$, the azimuthal symmetry along the $z$-direction allows us to find solutions to $g_z$ that are independent of $\phi$. Solving for $g_z(k,\theta)$, we find
\begin{equation}
	g_z(k,\theta)= \frac{\chi \zeta_k v_k}{i\omega+\eta}
	+\frac{1-\zeta_k^2}{1+\chi\zeta_k\cos\theta}\;
	\frac{v_k\cos\theta}{i\omega+1/\tau} \,, 
 \la{sol-g}
\end{equation}
where $\eta\to+0$ in the absence of chirality flipping and will be replaced by $1/\tau_{v}$ if the chirality flipping processes are taken into account.  


The phase space factor (\ref{upsilon}) takes into account the redistribution of the density of states along the Fermi surface in weak magnetic fields, i.e., accumulation of states at the south/north pole for left/right chirality, respectively (see Fig.~\ref{fig:FermiSurface}). However, it does not take into account the discreteness of Landau levels crucial for SdH oscillations. The discreteness of Landau level can be included through the Bohr-Sommerfeld quantization condition:
\begin{equation}
\frac{1}{2}\oint_{\gamma}\left(1+\frac{e}{\hbar}\bsym\Omega\cdot\bsym B\right) \bsym{\hat z}\cdot \bsym k\times\text d\bsym k=\frac{2\pi n}{\ell_B^{2}}\,. \la{BohrSom-LL}
\end{equation}

Eq. (\ref{BohrSom-LL}) comes from the non-trivial Poisson brackets between the coordinates of $\bsym k$ and $\gamma$ denotes the curves of constant energy. Taking into account the discreteness of Landau levels into Eq. (\ref{conducsimpl}), the conductivity per chirality becomes:
\begin{align}
\upsigma_{zz}^{(\chi)}=&-\frac{e^2}{2\pi^2}\sum_{n=0}^{\infty}\int\text dk \int\limits_{-1}^{1}\text d(\cos\theta)\,k^2\frac{\p f_0}{\p\varepsilon}v_k(\cos\theta+\chi\zeta_k)\nonumber
\\
&\times\delta\left(n-\frac{1-\cos^2\theta-2\chi\zeta_k\cos\theta}{4\zeta_k}\right)g_z(k,\theta)\,. \la{sigmazz}
\end{align}

Since the argument of the delta function has no real roots when $n\in \mathbb Z_-$, we can consider the sum starting from $n=-\infty$ and use the Poisson summation formula. Thus,
\begin{equation}
	\upsigma_{zz}^{(\chi)}=\upsigma_{zz}^{(0)}+2\sum_{l=1}^{\infty}\upsigma_{zz}^{(l)}
	\cos\left(\frac{\pi l}{2\zeta_F}+\frac{\pi}{4}\right)\,,
 \la{sigmaZZ}
\end{equation}
where we used dimensionless magnetic field
\be
	\zeta_{F}\equiv \zeta_{k}|_{k=k_{F}}=\frac{1}{2k_{F}^{2}l_{B}^{2}}=\frac{eB}{2\hbar k_{F}^{2}}\,.
 \la{zetaF}
\ee 

The non-oscillating part of (\ref{sigmaZZ}) is given by
\begin{align}
	\upsigma_{zz}^{(0)}=\frac{n_e e^2 v_F}{\hbar k_F}
	\left(\frac{1-\frac{12}{5}\zeta_F^2}{i\omega+1/\tau}
	+\frac{3\zeta_F^2}{i\omega+\eta}\right)\,,
 \la{sigmaZZ0}
\end{align}
where $n_e= k_F^3/(3\pi^2)$ is the total density of electrons per chirality. And, for the oscillating part we have
\begin{align}
	\upsigma_{zz}^{(l)}=\frac{n_e e^2 v_F}{\hbar k_F}
	\frac{1}{i\omega+1/\tau} 
	\frac{3}{2\pi}\frac{\lambda l}{\sinh\lambda l}\left(\frac{2\zeta_F}{l}\right)^{3/2}\,,
 \la{sigmaZZl}
\end{align}
where $\lambda = \pi^{2}T/(\hbar k_{F}v_{F}\zeta_{F})$. In the DC limit and in the absence of magnetic field, $\zeta_{F}=0$, Eqs. (\ref{sigmaZZ}-\ref{sigmaZZl}) are reduced to a standard Drude formula appropriately modified for Dirac spectrum:
\be
	\sigma_{0} = \frac{n_e e^2\tau}{\hbar k_F/v_{F}}\,.
\ee

In finite magnetic field the second term of (\ref{sigmaZZ0}) describes an ideal conductivity. In the absence of chirality flipping this conductivity diverges in static limit $\omega\to 0$. In more realistic models the process of chirality flipping are always present and one should replace $\eta\to 1/\tau_{v}$, where $\tau_{v}$ is a mean chirality lifetime. As the scattering with and without changes of chirality are due to very different processes one should expect the ratio $\tau_{v}/\tau$ to be significant.
Both $\tau$ and $\tau_v$ can in principle, be extracted from optical conductivity measurements. 



There are two small parameters in the regime of interest of this work. One is $\zeta_{F}$, i.e., the weakness of the magnetic field compared to the Fermi scale. The other is the smallness of temperature compared to the Fermi energy. We do not, however, make any assumptions on the relative size $\lambda$ of these small parameters. In deriving (\ref{sigmaZZ}-\ref{sigmaZZl}) we kept the leading ($B$-independent) and next to the leading terms of the expansion in $\zeta_{F}$ but restricted the expansion only to the leading term in $T/\varepsilon_{F}$. This is why the only temperature dependence in (\ref{sigmaZZ}-\ref{sigmaZZl}) is through the parameter $\lambda$. This means that we omitted all corrections proportional to $T/\varepsilon_{F}$ which could be comparable to the ones proportional to $\zeta_{F}$. The former corrections, however, are not universal and do not affect the magnetic field dependence of the conductivity.

A very convenient way to exclude the non-universal temperature corrections is to study the ratio $\sigma_{zz}(B)/\sigma_0$. In DC limit ($\omega\to 0$), it is given by
\bea
	\frac{\sigma_{zz}(B)}{\sigma_0}
	&=& 1 +3\left(\frac{\tau_{v}}{\tau}-\frac{4}{5}\right)\zeta_{F}^{2}+\frac{3}{\pi}
 \la{szz-ratio} \\
	&\times& \sum_{l=1}^{\infty}e^{-\lambda_D l}\frac{\lambda l}{\sinh\lambda l}\left(\frac{2\zeta_F}{l}\right)^{3/2}
	\cos\left(\frac{\pi l}{2\zeta_F}+\frac{\pi}{4}\right)\,. \la{cond-ratio}
 \nonumber
\eea

In the last equation we introduced the Dingle factor $\lambda_D=\pi\Gamma/(\hbar v_F k_F\zeta_F)$, which accounts for the smearing of LLs due to impurities.
In the case homogeneous sample  $\Gamma=\hbar/\tau_Q$ with the ``quantum time'' $\tau_{Q}$ determined by impurity scattering and equal to quasiparticles' lifetime.

If either $\lambda\gg 1$ or $\lambda_{D}\gg 1$, i.e., the temperature or smearing of Landau levels is larger than the gap between Landau levels, the oscillations in (\ref{szz-ratio}) disappear and the conductivity is given by the first line in (\ref{szz-ratio}). For smaller temperatures and Landau level smearing oscillations appear and become less and less harmonic with a further decrease of both $\lambda$ and $\lambda_{D}$.

\begin{figure}[h]
    \centering
    \includegraphics[width=0.47\textwidth]{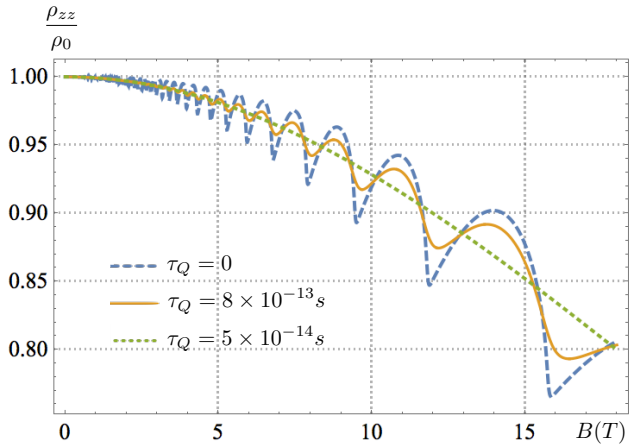}
    \caption{Longitudinal magnetoresistance as a function of the magnetic field. We used numerical values consistent with \cite{Ong2014}: $k_F=3.8\times10^8m^{-1}$, $v_F=9.3\times10^{5}m/s$,  $\tau=8\times10^{-13}s$, $T=2.5K$. The plots are made for three values of $\tau_{Q}/\tau=0, 1, 16$ and for $\tau_v=10\tau$. 
    }
    \label{fig:oscillations}
\end{figure}

In Fig.~\ref{fig:oscillations} we plot the magnetoresisitivity given by the inverse of expression in Eq. (\ref{szz-ratio}) for parameters consistent with the recent experiment on Cd$_3$As$_2$ \cite{Ong2014}. Comparison with the Fig.~4B of \cite{Ong2014} shows that the approach to magnetotransport in Dirac semimetals developed here describes qualitatively the emergence of quantum SdH oscillations and the tendency to negative magnetoresistance at strong magnetic fields (but still small $\zeta_{F}$) observed experimentally in Cd$_3$As$_2$ \cite{Ong2014}\footnote{The negative MR was not observed in \cite{Analytis-2015}.}. However the Cd$_3$As$_2$ data exhibit also a strong positive magnetoresistance present in weak magnetic fields \cite{Ong2014}. The more thorough comparison of our theory with experimental data requires an explanation of the positive magnetoresistance.  Since a (much weaker) positive MR has also been observed in the weak magnetic field region in Dirac semimetals ZrTe$_5$ and Na$_3$Bi, it is likely that this effect is generic for three-dimensional chiral materials. Possible explanations include the magnetic field dependence of impurity screening, the weak antilocalization and surface effects resulting from Fermi arcs \cite{Analytis-2015}. We leave the systematic treatment of these effects, as well as the study of microscopic mechanisms of chirality relaxation, for future studies. 

We would like to thank I. Aleiner for useful discussions. The work was supported in part by the NSF under grant no. DMR-1206790 (A.G.A), and by the U.S. Department of Energy under Contracts DE-FG-88ER40388 and DE-SC-0012704 (D.K.).





\onecolumngrid
\newpage
\twocolumngrid
\setcounter{equation}{0}

\renewcommand{\theequation}{A.\arabic{equation}}

\section{Appendix A: Boltzmann Equation}

In this section, we will derive the expression (\ref{sol-g}) by solving the stationary Boltzman equation. Rewriting Eq. (\ref{Boltz-g}) in spherical coordinates and plugging the formula for scattering rate (\ref{scat-rate}) into it, we obtain:
\begin{align}
&\left(i\omega\Upsilon+2k\zeta_k v_k\frac{\p}{\p\phi}\right)\bsym g(\bsym k)-v_k\bsym{\hat k}= \la{Boltz-linear}
\\
&=v_k\chi\zeta_k\bsym{\hat z}+\frac{3\Upsilon}{16\pi^3}\int\text d^3k'\Upsilon'(\bsym g'-\bsym g)\frac{(1+\boldsymbol{\hat k'}\cdot \boldsymbol{\hat k})}{\nu(\varepsilon)\tau(\varepsilon)}\delta(\varepsilon-\varepsilon').\nonumber 
\end{align}

Integrating Eq. (\ref{Boltz-linear}) over the solid angle, we obtain:
\begin{equation}
\int_{\mathbb S^2}\text d\phi\,\text d(\cos\theta)\,\Upsilon(\bsym k)\bsym g(\bsym k)=\frac{4\pi\chi\zeta_k\bsym{\hat z}}{i\omega}. \la{constraint}
\end{equation}

It is obvious from here that there are no stationary solutions for Eq. (\ref{Boltz-g}) when $\omega\rightarrow0$. The azimuthal symmetry allows to find solutions of (\ref{Boltz-linear}) that are independent of $\phi$ for $g_z$. After the integration over $(k',\phi')$, we end up with:
\begin{align}
i\omega\Upsilon g_z=\;&v_k(\cos\theta+\chi\zeta_k)+\int_{-1}^{1}\text d(\cos\theta')\Upsilon'(g'_z-g_z)\Upsilon\nonumber
\\
&\times\frac{3}{4\tau}(1+\cos\theta\cos\theta')\,.
\end{align}

The easiest way to solve this equation is to expand $\Upsilon g_z$ in terms of Legendre polynomials and use their orthogonality conditions. Thus,
\begin{align}
\Upsilon g_z&=\sum_{l=0}^{\infty}(2l+1)a_l(k)P_l(\cos\theta),\nonumber
\\
&=\frac{\chi\zeta_kv_k}{i\omega}+\left[\frac{\zeta_k^2}{i\omega}+\frac{\left(1-\zeta_k^2\right)}{i\omega+\tau^{-1}}\right]v_k\cos\theta,
\end{align}
where $a_0$ is obtained through (\ref{constraint}).

\setcounter{equation}{0}

\renewcommand{\theequation}{B.\arabic{equation}}
\section{Appendix B: Discreteness of Landau levels}

Quantum effects in the conductivity can be implemented through the Bohr-Sommerfeld quantization condition. The prescription here is the same one used in the old quantum theory; given a classical system, we introduce quantum effects by imposing that canonical variables satisfy:
\[
\oint_\gamma p_i\,\text dq_i=2\pi\hbar\,(n_i+\tfrac{1}{4}\text{ind} \gamma),
\]
where $\gamma$ is a curve in phase space in which the Hamiltonian is a constant and $\text{ind} \gamma$ is the Maslov index of $\gamma$. 

However, in the presence of a nonvanishing Berry curvature, the perpendicular components of $\bsym k$\footnote{With respect to $\bsym B$.} fail to be canonically conjugated. Instead,
\begin{equation}
\frac{1}{2}\varepsilon^{abc}\left(1+\frac{e}{\hbar}\bsym\Omega\cdot\bsym B\right)\{k_b,k_c\}=\frac{eB^a}{\hbar^2}\,.
\end{equation}

Following the same recipe and using that $\bsym B=B\bsym{\hat z}$, the discreteness of Landau levels can be imposed by assuming that:
\begin{equation}
\frac{1}{2}\oint_{\gamma}\left(1+\frac{e}{\hbar}\bsym\Omega\cdot\bsym B\right) \bsym{\hat z}\cdot \bsym k\times\text d\bsym k=\frac{2\pi}{\ell_B^{2}}(n+\tfrac{1}{4}\text{ind} \gamma)\,. \la{BohrSom}
\end{equation}

It implies the area quantization -- in units of $2\pi/\ell_B^{2}$ -- for the section of the Brillouin zone with $k_z$ constant. We can find the surfaces with constant $n$ in $k$-space by solving equation (\ref{BohrSom}):
\begin{align}
n+\frac{1}{4}\text{ind} \gamma&=\frac{\ell_B^2}{2}\left(k_\perp^2-\frac{\chi k_z}{\ell_B^2\sqrt{k_\perp^2+k_z^2}}\right),
\\
&=\frac{1}{4\zeta_k}\left(1-\cos^2\theta-2\chi\zeta_k\cos\theta\right). \label{n-eq}
\end{align}

If we impose that $n=0$ is the smallest possible integer solution of (\ref{n-eq}) and use the fact that $\text{ind} \gamma\in \mathbb Z$; the only possible values of the Maslov index are $\{-2,-1,0\}$.

\setcounter{equation}{0}

\renewcommand{\theequation}{C.\arabic{equation}}
\section{Appendix C: SdH oscillations}

In this section, we will apply the Bohr-Sommerfeld quantization prescription to introduce quantum effects in the conductivity. Assuming that the only contribution to transport comes from the discrete levels, the conductivity per chirality becomes:
\begin{align}
\upsigma_{zz}^{(\chi)}&=-\frac{e^2}{2\pi^2}\sum_{n=-\infty}^{\infty}\int\text dk \int\limits_{-1}^{1}\text d(\cos\theta)\,k^2\frac{\p f_0}{\p\varepsilon}(\cos\theta+\chi\zeta_k)\nonumber
\\
&\times \delta[n-\tfrac{1}{4}\left(\sin^2\theta/\zeta_k-2\chi\cos\theta-\text{ind}\gamma\right)]v_k\, g_z,
\end{align}
where $g_z$ is given in (\ref{sol-g}). We have used that there is no real solution for (\ref{n-eq}) when $n\in\mathbb Z_-$. Therefore, all surfaces for negative integer $n$ are outside of the integration range. The integral over $k$ is performed near Fermi surface. 

Using the Poisson formula,
\[
\sum_{n=-\infty}^{\infty}\delta(x-n)=\sum_{l=-\infty}^{\infty}e^{i2\pi ln},
\]
the conductivity can be rewritten as:
\begin{equation}
\upsigma_{zz}^{(\chi)}=-\frac{e^2}{2\pi^2}\sum_{l=-\infty}^{\infty}\left(\frac{\mathcal I_1^{(l)}}{i\omega+\tau_v^{-1}}+\frac{\mathcal I_2^{(l)}}{i\omega+\tau^{-1}}\right),
\end{equation}
where,
\begin{align}
\mathcal I_1^{(l)}=&\int \text d k\,k^2v_k^2\,\frac{\p f_0}{\p\varepsilon}\,\chi\zeta_k\,e^{i\frac{\pi l}{2}(\zeta_k^{-1}+\zeta_k-\text{ind}\gamma)} \la{I1}
\\
&\times\int\limits_{-1}^{1}\text d(\cos\theta)(\cos\theta+\chi\zeta_k)\,e^{-i\frac{\pi l}{2}(\cos\theta+\chi\zeta_k)^2}, \nonumber
\\
\mathcal I_2^{(l)}=&\int \text d k\,k^2v_k^2\,\frac{\p f_0}{\p\varepsilon}\,(1-\zeta_k^2)\,e^{i\frac{\pi l}{2}(\zeta_k^{-1}+\zeta_k-\text{ind}\gamma)} \la{I2}
\\
&\times\int\limits_{-1}^{1}\text d(\cos\theta)\frac{\cos\theta+\chi\zeta_k}{1+\chi\zeta_k\cos\theta}\,\cos\theta\,e^{-i\frac{\pi l}{2}(\cos\theta+\chi\zeta_k)^2}.\nonumber
\end{align}

The integral in Eq. (\ref{I1}) accounts for the intervalley scattering and can be easily calculated:
\begin{equation}
\mathcal I_1^{(l)}=-\left(\frac{2}{\hbar}v_Fk_F^2\zeta_F^2\right)\delta_{l,0}\,.
\end{equation}

In the equation above, we have used that 
\[
\frac{\p f_0}{\p\varepsilon}\approx-\frac{\delta(k-k_F)}{\hbar v_F}.
\] 

The terms with $l\neq 0$ vanish since the chirality relaxation mechanism that we have considered only accounts for the scattering between the zero-modes. 

Let us now consider the contribution for the intravalley scattering coming from Eq. (\ref{I2}). For $l=0$ the integral can be performed analytically, however, we are only interested in the range where the semiclassical picture is valid. If we restrict ourselves terms up to $\mathcal O(\zeta_F^2)$, we end up with:
\begin{equation}
\mathcal I_2^{(0)}=-\frac{2}{\hbar}v_Fk_F^2\left[\frac{1}{3}-\frac{7}{15}\zeta_F^2+\mathcal O(\zeta_F^4)\right].
\end{equation}

In order to calculate $\mathcal I_2^{(l)}$ for $l\neq0$, it is convenient to define $x=\cos\theta+\chi\zeta_k$. Expanding the integrand up to $\mathcal O(\zeta_k^2)$, we find that:
\begin{align}
\mathcal I_2^{(l)}&=\int \text d k\,k^2\,v_k^2\,\frac{\p f_0}{\p\varepsilon}\,\exp\left[i\frac{\pi l}{2}\left(\zeta_k-\text{ind}\gamma-1/\zeta_k\right)\right]\nonumber
\\
&\times[-\chi\zeta_k\mathcal Q_1+(1+\zeta_k^2)\mathcal Q_2-\zeta_k\chi\mathcal Q_3+\zeta_k^2\mathcal Q_4], \la{I2-l}
\end{align}
where 
\begin{equation}
\mathcal Q_m\equiv\int\limits_{-1+\chi\zeta_k}^{1+\chi\zeta_k}\text dx \,x^m\,\exp\left(-i\frac{\pi l}{2\zeta_k}x^2\right).
\end{equation}

Solving for odd values of $m$:
\begin{align*}
\mathcal Q_1&=\frac{2\zeta_k}{\pi l}\,e^{-i\frac{\pi l}{2\zeta_k}(\zeta_k^{-1}+\zeta_k)}\sin(\pi l\chi)=0,
\\
\mathcal Q_3&=\frac{4\chi i(-1)^l\zeta_k^2}{\pi l}\,e^{-i\frac{\pi l}{2}(\zeta_k^{-1}+\zeta_k)}.
\end{align*}

However, $\zeta_k\mathcal Q_3= \mathcal O(\zeta_k^3)$ and such term can be neglected. Let us now focus on $m$ even. They can all be obtained through $\mathcal Q_0$ as follows:
\[
\mathcal Q_m=\left(\frac{2i\zeta_k}{\pi}\right)^{m/2}\frac{d^{m/2}}{dl^{m/2}}\mathcal Q_0\,,
\]
where $l$ is set to be a non-zero integer at the end of the calculation. Clearly, $\mathcal Q_4= \mathcal O(\zeta_k^3)$ and we only need to calculate $\mathcal Q_2$. The integral $\mathcal Q_0$ can be written as:
\begin{equation}
\mathcal Q_0=\int\limits_{0}^{1+\chi\zeta_k}\text dx\,e^{-i\frac{\pi l}{2\zeta_k}x^2}+\int\limits_{0}^{1-\chi\zeta_k}\text dx\,e^{-i\frac{\pi l}{2\zeta_k}x^2}. \la{Fresnel}
\end{equation}

Let us focus on the right hand side of Eq. (\ref{Fresnel}). Thus,
\begin{equation*}
\int\limits_{0}^{1\pm\chi\zeta_k}\text dx \,e^{-i\frac{\pi l}{2\zeta_k} x^2}=\frac{1}{2}\sqrt{\frac{2\zeta_k}{il}}-\frac{1}{2}\sqrt{\frac{2\zeta_k}{\pi l}}F_0\left(\tfrac{\pi l(1\pm\chi\zeta_k)^2}{2\zeta_k}\right),
\end{equation*}
where we have defined: 
\[
F_m(t)=\int_t^\infty\text dy\,y^{-m-1/2}e^{-iy}\,.
\] 

After integration by parts, one can show that:
\[
F_m(t)=-i\frac{e^{-it}}{t^{m+1/2}}+i\left(m+\frac{1}{2}\right)F_{m+1}(t).
\]

Since we are restricting ourselves to terms up to $\mathcal O(\zeta_k^{2})$, we obtain:
\begin{align*}
\mathcal Q_0=&\;\sqrt{\frac{2\zeta_k}{il}}+\frac{2\zeta_k (-1)^l}{\pi l}\frac{e^{-i\frac{\pi l}{2}(\zeta_k^{-1}+\zeta_k)}}{1-\zeta_k^2}\,,
\\
\mathcal Q_2=&-\frac{e^{\frac{i\pi}{4}}}{\pi}\sqrt{\frac{2\zeta_k^3}{l^3}}+\frac{2\zeta_k (-1)^l}{\pi l}e^{-i\frac{\pi l}{2}(\zeta_k^{-1}+\zeta_k)}\left(i+\frac{2\zeta_k}{\pi l}\right).
\end{align*}

Plugging all determined values for $\mathcal Q_m$ into (\ref{I2-l}):
\begin{align}
\mathcal I_2^{(l)}=&-\frac{2 k_F^2 v_F\zeta_F}{\hbar\pi l}e^{i\pi l(1-\frac{1}{2}\text{ind}\gamma)}\left(i+\frac{2\zeta_F}{\pi l}\right) \la{I2-int}
\\
&-\frac{\sqrt{2}e^{i\frac{\pi}{2}(\frac{1}{2}-\text{ind}\gamma\, l)}}{\hbar\pi l^{3/2}}\int\text d\varepsilon\,k^2v_k\zeta_k^{3/2}\frac{\p f_0}{\p\varepsilon}e^{i\frac{\pi l}{2}(\zeta_k^{-1}+\zeta_k)}\,.\nonumber
\end{align} 

Here we have used that we can invert the dispersion relation and write $k(\varepsilon)$. The energy integral is performed at the vicinity of the Fermi surface. Since we assume that $T/\varepsilon_F\ll 1$, all the integrand besides the oscillating exponential is consider to vary slowly in the temperature range. In addition to that, we must expand the exponent near the Fermi energy. Keeping only linear deviations in the exponent, we are left with:
\begin{align}
\mathcal I_2^{(l)}&\approx-\frac{2 k_F^2 v_F\zeta_F}{\hbar\pi l}e^{i\pi l(1-\frac{1}{2}\text{ind}\gamma)}\left(i+\frac{2\zeta_F}{\pi l}\right) \la{I2-T}
\\
&-\frac{e^{i\frac{\pi}{2}[\frac{1}{2}+l(\zeta_F^{-1}-\text{ind}\gamma)]}}{2\hbar\pi}k_F^2v_F\left(\frac{2\zeta_F}{l}\right)^{3/2} \int\limits_{-\infty}^{\infty}\text dt\frac{e^{(1+i\lambda l/\pi)t}}{(e^t+1)^2}.\nonumber
\end{align} 

In the Eq. (\ref{I2-T}), we have defined $t=(\varepsilon-\mu)/T$ and $\lambda =\pi^2T/(\hbar v_F k_F\zeta_F)$. The integral can be solved using the residue theorem, and its value is given by:
\[
 \int\limits_{-\infty}^{\infty}\frac{e^{(1+i\lambda l/\pi)t}}{(e^t+1)^2}\text dt=\frac{\lambda l}{\sinh\lambda l}.
\]

Therefore, the conductivity can be expressed as:
\begin{equation}
	\upsigma_{zz}^{(\chi)}=\upsigma_{zz}^{(0)}+2\sum_{l=1}^{\infty}\upsigma_{zz}^{(l)}
	\cos\left[\frac{\pi l}{2}\left(\frac{1}{\zeta_F}-\text{ind}\gamma\right)+\frac{\pi}{4}\right], \la{cond-ind}
\end{equation}
where
\begin{align}
\upsigma_{zz}^{(0)}=&\frac{n_e e^2 v_F}{\hbar k_F}\left[\frac{3\zeta_F^2}{i\omega+\tau_v^{-1}}+\frac{1}{i\omega+\tau^{-1}}\right. \la{sig0}
\\
&\times\left.\left(1+\frac{3}{2}\zeta_F\,\delta_{1+\text{ind}\gamma,0}-\frac{12}{5}\zeta_F^2+\frac{3}{4}(\text{ind}\gamma)^2\zeta_F^2\right)\right],\nonumber
\end{align}
and
\begin{align}
\upsigma_{zz}^{(l)}=\frac{n_e e^2 v_F}{\hbar k_F}
	\frac{1}{i\omega+1/\tau} 
	\frac{3}{2\pi}\frac{\lambda l}{\sinh\lambda l}\left(\frac{2\zeta_F}{l}\right)^{3/2}. \la{sig-l}
\end{align}

In the Eq. (\ref{sig0}), we have used that:
\[
\frac{12}{\pi^2}\sum_{l=1}^{\infty}(-1)^l\frac{\cos(\tfrac{1}{2}l\pi\,\text{ind}\gamma)}{l^2}=\frac{3}{4}(\text{ind}\gamma)^2-1.
\]

Although $\text{ind}\gamma$ can in principle be obtained by the WKB calculation, we assume $\text{ind}\gamma=0$ in the main text.

\setcounter{equation}{0}

\renewcommand{\theequation}{D.\arabic{equation}}
\section{Appendix D: Dingle factor}

In the treatment of quantum oscillations, the Dingle factor in Eq. (\ref{cond-ratio}) comes from the smearing of LLs due to impurity scattering. In the previous section, we assumed that the density of states have sharp peaks at each Landau level. However, this is not true in a more realistic scenario. The presence of impurities breaks the energy degeneracy of the Landau levels and as a net result they get smeared by the presence of impurities. 

The assumption that $k_F\ell_B^2/(v_F\tau)\ll 1$, allows us to disregard corrections to the plane-wave scattering due to the magnetic field\footnote{Otherwise, we must consider the whole matrix elements of the impurity potential in the presence of magnetic field.}. Within this approximation, the density of states is still isotropic, however, it gets a contribution coming from the smearing of energy levels, namely:
\begin{align}
\nu(\xi)&=\int_{\mathbb{BZ}}\text d^3k\,\Im\left[G^R(\xi,\bsym k)\right],\nonumber
\\
\nu(\xi)&=\frac{1}{\pi}\int_{\mathbb{BZ}}\text d^3k\,\frac{\Gamma(\xi, k)}{[\xi-\varepsilon(k)]^2+\Gamma^2(\xi, k)}.
\end{align}

In the limit when $\Gamma\rightarrow0$, we recover the well-know result 
\[
\nu(\xi)=\int_{\mathbb{BZ}}\text d^3k\;\delta(\xi-\varepsilon(k))=\left.\frac{4\pi k^2}{\hbar v_k}\right|_{\varepsilon=\xi}.
\]

In fact, the smearing of the energy levels can be introduced by following replacement:
\[
\delta(\xi-\varepsilon(k))\rightarrow\frac{1}{\pi}\frac{\Gamma(\xi, k)}{[\xi-\varepsilon(k)]^2+\Gamma^2(\xi, k)}.
\]

We can thus rewrite Eq. (\ref{I2-int}) in a more convenient way:
\begin{align}
\mathcal I_2^{(l)}&=-\frac{2 k_F^2 v_F\zeta_F}{\hbar\pi l}(-1)^l\left(i+\frac{2\zeta_F}{\pi l}\right)-\frac{\sqrt{2}e^{i\frac{\pi}{4}}}{\pi l^{3/2}} 
\\
&\times\int\text d\xi\int\limits_{\frac{1}{\sqrt2\ell_B}}^{\infty}\text dk\,\delta(\xi-\varepsilon(k))\,k^2v_k^2\,\zeta_k^{3/2}\frac{\p f_0}{\p\xi}e^{i\frac{\pi l}{2}(\zeta_k^{-1}+\zeta_k)}\,.\nonumber
\end{align} 

Here, we have set $\text{ind}\gamma$ to zero in order to shorten up the notation since this factor brings no extra difficulties. The choice of the lower limit of integration is for later convenience. As previously mentioned, the smearing can be taken into account by replacing the delta function in the integral above by a Lorentzian distribution. Therefore, let us focus on:
\begin{equation}
\mathcal S_l(\xi)=\frac{1}{\pi}\int\limits_{\frac{1}{\sqrt2\ell_B}}^{\infty}\text dk\,\frac{\Gamma(\xi, k)\,e^{i\frac{\pi l}{2}(\zeta_k^{-1}+\zeta_k)}}{[\xi-\varepsilon(k)]^2+\Gamma^2(\xi, k)}k^2v_k^2\zeta_k^{3/2}. \la{S-Din}
\end{equation}

The cutoff $1/(\sqrt2\ell_B)$ guarantees the integral convergence. One can solve Eq. (\ref{S-Din}) using the steepest descent approximation. For that, let us analytically continue the integrand and define $z\equiv \zeta_k^{-1/2}$. Hence,
\begin{equation*}
\mathcal S_l(\xi)=\frac{1}{\sqrt{8}\pi\ell_B^3}\int\limits_{1}^{\infty}\text dz\, H\left(\xi,\tfrac{z}{\sqrt2\ell_B}\right)\frac{e^{i\frac{\pi l}{2}(z^2+1/z^2)}}{z},
\end{equation*}
where 
\[
H(\xi,k)\equiv\frac{\Gamma(\xi,k)\,v_k^2}{[\xi-\varepsilon(k)]^2+\Gamma^2(\xi,k)}.
\]

Expanding the exponent near $z=1$, we obtain:
\begin{align}
\mathcal S_l(\xi)&\approx \frac{(-1)^l}{\sqrt{8}\pi\ell_B^3}\,H\left(\xi,\tfrac{1}{\sqrt2\ell_B}\right)\int_{\mathcal C}\text dz\,e^{i2\pi l(z-1)^2}
\\
&+\frac{i}{\sqrt2\ell_B^3}\,\text{Res}_{\varepsilon\rightarrow\xi+i\Gamma}\left[H\left(\xi,\tfrac{z}{\sqrt2\ell_B}\right)\frac{e^{i\frac{\pi l}{2}(z^2+1/z^2)}}{z}\right].\nonumber
\end{align}

The contour $\mathcal C$ is defined by $\Re[(z-1)^2]=0$ together with $\Im[(z-1)^2]\geq0$ and $|z|\geq 1$. Let us assume for simplicity that $\Gamma(\xi,k)=\Gamma(\xi)$. Using that only $\xi\sim\mu$ contributes to $\mathcal I_2^{(l)}$, we find that:
\begin{equation}
\mathcal S_l(\xi)\approx\frac{(-1)^l v_0^2}{8\pi\ell_B^3\sqrt{2l}}\frac{\Gamma}{\xi^2}+\frac{k^2v_k\zeta_k^{3/2}}{\hbar}\exp\left(\frac{i\pi l}{2\zeta_k}\right),
\end{equation}
where 
\[
v_0\equiv v_k|_{k=\tfrac{1}{\sqrt2\ell_B}},
\]
and $k$ is taken to be $k(\xi+i\Gamma)$. Plugging it into $\mathcal I_2^{(l)}$, we end up with:
\begin{align}
\mathcal I_2^{(l)}&=-\frac{2 k_F^2 v_F\zeta_F}{\hbar\pi l}(-1)^l\left(i+\frac{2\zeta_F}{\pi l}\right)+\frac{k_F^3v_0^2\Gamma\zeta_F^{3/2}e^{i\frac{\pi}{4}}}{\pi^2 l^2\varepsilon_F^2}\nonumber
\\
&-\frac{e^{i\frac{\pi l}{2\zeta_F}+i\frac{\pi}{4}-\lambda_D l}}{2\hbar\pi}k_F^2v_F\left(\frac{2\zeta_F}{l}\right)^{3/2} \int\limits_{-\infty}^{\infty}\text dt\,\frac{e^{(1+i\lambda l/\pi)t}}{(e^t+1)^2}.\nonumber
\end{align} 

Here, we have defined $\lambda_D=\pi\Gamma/(\hbar k_F v_F\zeta_F)$ and neglected terms of $\mathcal O(\Gamma e^{-\lambda_D l})$. However, from (\ref{scales}),
\[
\Gamma/\varepsilon_F\sim\Gamma/(\hbar v_Fk_F)\ll\zeta_F
\]
and consequently $\zeta_F^{3/2}\Gamma/\varepsilon\ll\zeta_F^2$. Therefore, the second term in $\mathcal I_2^{(l)}$ can also be neglected within our approximation. 

The only modification in the  conductivity expression coming from the smearing of LLs occurs in Eq. (\ref{sig-l}), which must be replaced by:
\begin{align}
\upsigma_{zz}^{(l)}=\frac{n_e e^2 v_F}{\hbar k_F}
	\frac{e^{-\lambda_Dl}}{i\omega+1/\tau} 
	\frac{3}{2\pi}\frac{\lambda l}{\sinh\lambda l}\left(\frac{2\zeta_F}{l}\right)^{3/2}. \la{sig-l}
\end{align}

\end{document}